\documentclass[twocolumn,showpacs,preprintnumbers,amssymb,prl]{revtex4}
\usepackage{graphicx}
\usepackage{dcolumn}
\usepackage{color}
\usepackage{bm}

\def \be{\begin{equation}}
\def \ee{\end{equation}}
\def \ben{\begin{eqnarray}}
\def \een{\end{eqnarray}}

\begin{document}

\title{On A Cosmological Invariant as an Observational Probe in the Early Universe }
 
\author{Debashis Gangopadhyay}
\altaffiliation{debashis@bose.res.in}
\affiliation{S. N. Bose National Centre for Basic Sciences,
JD Block, Sector-III, Salt Lake, Kolkata 700 098, India.}

\author{Somnath Mukherjee}
\altaffiliation{sompresi@gmail.com}
\affiliation{Dharsa Mihirlal Khan Institution[H.S],
            P.O:-New G.I.P Colony,  Dist:-Howrah-711112, India. }

\begin{abstract}

k-essence scalar field models are usuually taken to have lagrangians of the form 
${\mathcal L}=-V(\phi)F(X)$
with $F$ some general function of $X=\nabla_{\mu}\phi\nabla^{\mu}\phi$. Under certain conditions  this
lagrangian in the  context of the early universe can take the form of that 
of an oscillator with time dependent frequency. The Ermakov  invariant for a time dependent oscillator  
in a cosmological scenario then leads to an invariant quadratic form involving the Hubble parameter 
and the logarithm of the scale factor. In principle, this invariant can lead to further observational 
probes for the early universe. Moreover, if such an invariant can be observationally verified then 
the presence of dark energy will also be indirectly confirmed.
 
\end{abstract}


\maketitle

{\bf 1.Introduction}

The motivation for this work lies in the existence of 
an invariant related to the time dependent oscillator,
first obtained by Ermakov \cite{ermakov},\cite{lewis},\cite{milne}.
In the context of the $k-$essence lagrangian \cite{gango1},
the logarithm of the scale factor
in a homogeneous universe at very early times after the big bang
satisfies the equations of motion of an
oscillator with time-dependent frequency \cite{gango2}. 
The classical solutions of this theory are fully consistent with the
inflationary scenario and a radiation dominated universe. A  measure of temperature
fluctuations can also be estimated using standard prescriptions \cite{gango2}.
In this work the focus will be on another interesting aspect of the time dependent
oscillator ,{\it viz.}, the existence of invariants or first integrals of motion
\cite{ermakov}. Here we show that as the $k-$ essence lagrangian takes the form of that 
of a time dependent oscillator,the invariant has cosmological 
analogues--in the classical as well as a quantum context. 
Classically one can construct an invariant quadratic form involving the Hubble parameter
and the logarithm of the scale factor. Quantum expectation values of a function containing 
the scale factor and Hubble parameter can also be obtained. The quantum aspects will be 
discussed in subsequent publications.In this work we will limit ourselves to the classical aspects.
Existence of this invariant implies possibilities of 
further observational probes in the early universe.

First a brief review is in order.
Scherrer \cite{scherrer} showed that it is possible to unify the dark matter and dark energy
components into a single scalar field model with the scalar field $\phi$ having a
non-canonical kinetic term. These scalar fields are the $k-$essence fields
which first appeared in models of inflation \cite{armendariz} and subsequently
led to models of dark energy also \cite{chiba}.
The general form of such lagrangians is 
some function $F(X)$ with $X=\nabla_{\mu}\phi\nabla^{\mu}\phi$, and do not depend explicitly
on $\phi$ to start with. In \cite{scherrer}, a 
scaling relation was obtained ,{\it viz.} $X ({dF\over dX})^{2}=Ca(t)^{-6}$, $C$ a
constant (similar expression was also derived in \cite{chimento1}).
\cite{gango1} incorporates the scaling relation
of \cite{scherrer} and in \cite{gango2} it is shown how this lagrangian can be 
approximated to that of a time dependent oscillator for small scale factors in a 
certain epoch of the early universe.
Literature on dark matter, dark energy and $k-$ essence
can be found in \cite{sahni}.

The lagrangian $L$ (or the pressure $p$) is taken as 
\ben
{\mathcal L}= -V(\phi) F(X)
\label{lag1}
\een
The energy density is 
\ben
\rho = V(\phi)[ F(X) -2 X F_{X}]
\label{energydensity}
\een
with $F_{X}\equiv {dF\over dX}$. 
In this work, the scalar potential $V(\phi)=V$ is a ({\it positive}) constant and all the
time variables $t\equiv t/t_{0}$, where $t_{0}$ is the present epoch and we are
interested only in $t< 1$ scenarios. Also $a(t_{1})< a(t_{2})$ for $t_{1}< t_{2}$ etc.

Using the scaling law and the zero-zero component of Einstein's field 
equations an expression for the lagrangian is obtained as follows.Take the 
Robertson-Walker (RW) metric :
$ds^{2}= c^{2}dt^{2} - a^{2}(t)[{dr^{2}\over (1-kr^{2})} 
+ r^{2}(d\theta^{2} + sin^{2}\theta d\phi^{2})]$.
where $k(=0, 1\enskip or -1)$ is the curvature constant.
The zero-zero component of Einstein's equation reads:
$R_{00} - {1\over 2}g_{00}R = - \kappa T_{00}$.
This gives with the RW metric 
${k\over a^{2}} + H^{2} = {8\pi G\over 3}\rho$.
For $k=0$, and a homogeneous and isotropic spacetime (i.e. $\phi(t,\bf x)=\phi(t)$)
(\ref{lag1}) becomes
\ben
{\mathcal L}=-c_{1}\dot q^{2} - c_{2}\dot \phi e^{-3q}
\label{lag2}
\een
with $q(t)=ln\enskip a(t)$, $c_{1}= 3(8\pi G)^{-1}$, $c_{2}=2V\sqrt C$,
and two generalised coordinates $q(t)$ and $\phi(t)$.
(\ref{lag2}) has a kinetic term for $q$ and an interaction term.
There is no kinetic term for $\phi$.

\includegraphics[angle=-90,width=8cm]{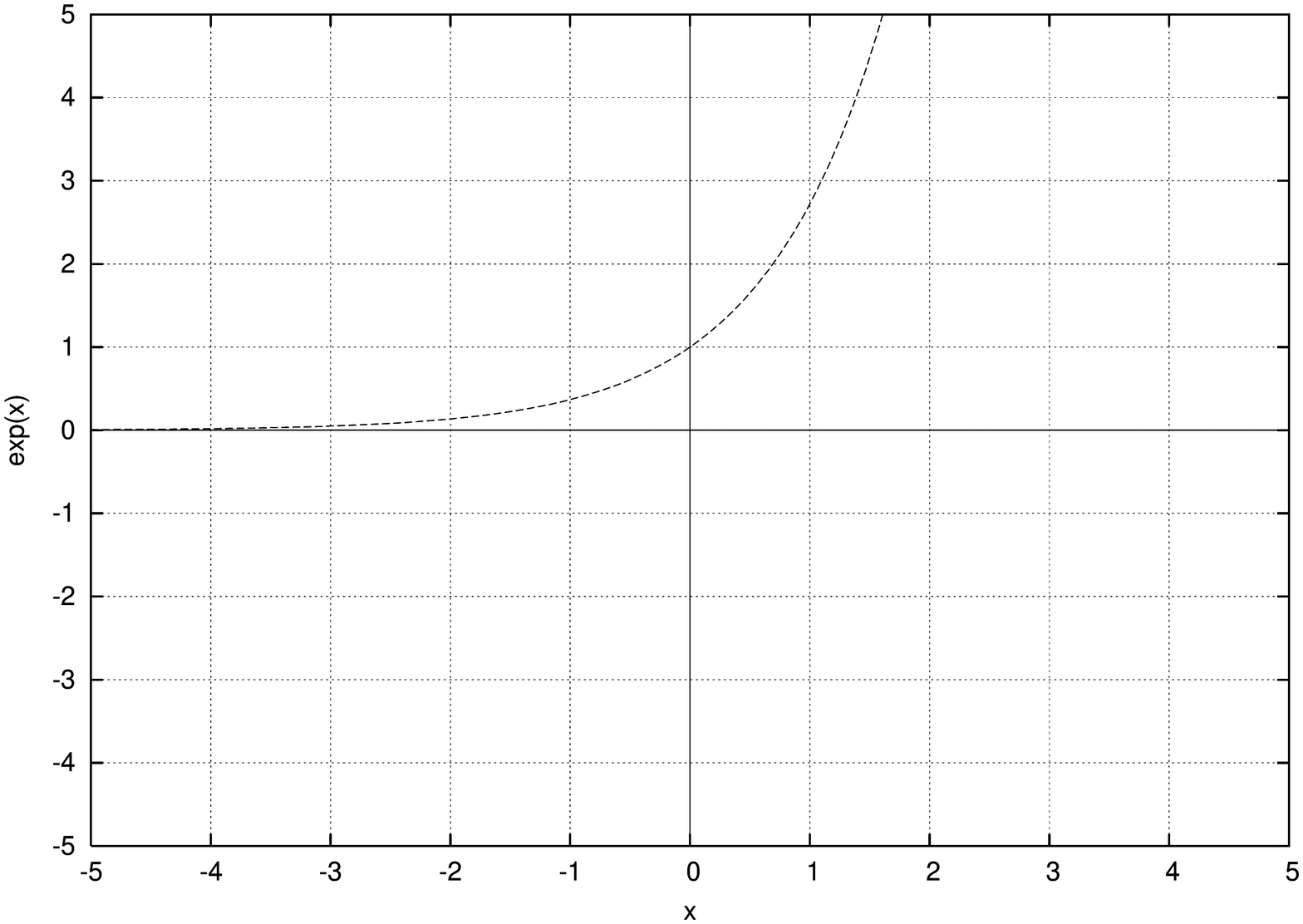}

Note that $a=e^{q}$. It is readily seen from the graph (in the figure $x\equiv q$) 
of the exponential function that in the region  $-1 < q < 0$ one has , $a=e^{q} < 1$ . 
Hence in this region $q$ small (i.e. $\vert q\vert <1$) means $a$ is also small (i.e $a < 1$). 
Moreover, in this region $a$ grows from $e^{-1}=0.367879$ to $e^{0}=1$. 
So within this region $a$ grows as $q$ grows. So smaller values of $q$ mean that we 
are going back to smaller values of $a$ i.e. to earlier epochs. 
{\it Throughout this work we will restrict ourselves to this domain i.e. $-1 < q < 0$. 
In this domain $a$ is small when $q$ is small and $|a|<1$ for $|q|<1$}. So expand the exponential in (\ref{lag2}),
keep terms upto $O(q^{2})$ and replace $q$ by $q+ {1\over 3}$ to get \cite{gango2}
\ben
{\mathcal L} = -{M\over 2}[\dot q ^{2} + 12\pi G g(t) q^{2}] -({1\over 2})g(t)
\label{lag3}
\een
where $M={3\over 4\pi G} $,
$g(t)= 2{\sqrt C}V\dot\phi$, 
and we use $c = 1$ ($c$ is speed of light).
Put  $12\pi G g(t) = -\Omega^{2}(t)$.
This means
\ben
\phi(t)= -{1\over 24\pi G {\sqrt C}V}\int dt \Omega ^{2}(t)
\label{phi1}
\een
(\ref{lag3}) now becomes
\ben
{\mathcal L} =-{M\over 2}[\dot q ^{2} - \Omega^{2} (t) q^{2}] -({1\over 2})g(t)
\label{lag4}
\een
Now, the term ${1\over 2}g(t)$ is a total time derivative
and thus has no contribution to the equations of motion and hence ignorable.
Then(\ref{lag4}) becomes
\ben
{\mathcal L} = -{M\over 2}[\dot q ^{2} - \Omega^{2} (t) q^{2}]
\label{lag5}
\een
Ignoring the overall negative sign in all subsequent discussions, we have a time dependent 
oscillator for  $q(t)=ln~a(t)$. 

{\bf 2. Ermakov Invariant in a cosmological context}

The Hamiltonian ${\mathcal H}$  corresponding to (\ref{lag5}) is 
\ben
{\mathcal H} = {M\over 2}[p^{2} + \Omega^{2}(t)q^{2}]
={M\over2}[H^{2} + \Omega^{2}(t) q^{2}]
\label{ham}
\een
where $p=\dot q= {\dot a\over a}= H$, $H$ is the Hubble parameter.
Following Ermakov \cite{ermakov}, \cite{lewis} , one can immediately 
write down the invariant $I$ as  
\ben
I = {1\over 2}[ \rho^{-2}q^{2} + \biggl(\rho H - {1\over M}\dot\rho q\biggr)^{2}]
\label{inv1}
\een
where $\rho(t)$ satisfies Ermakov's equation
\ben
{1\over M^{2}} \ddot\rho + \Omega^{2}\rho -\rho^{-3} =0
\label{erma}
\een
Putting in the values of $M$ and simplifying , one gets 
\ben
I= {\mathcal A(t)} (ln~a(t))^{2} -{\mathcal B(t)} (ln~a(t)) H(t) + {\mathcal C(t)} H^{2}\nonumber\\
\label{inv2}
\een
with
\ben
{\mathcal A(t)}= {\rho^{-2}(t)\over 2} + {32\pi^{2}G^{2}\over 9}(\dot\rho (t))^{2},\nonumber\\
{\mathcal B(t)} = {8\pi G\rho (t)\dot\rho (t)\over 3},\nonumber\\
{\mathcal C(t)} = {\rho^{2}(t)\over 2}
\label{functro}
\een
$I$ is an invariant for the Hamiltonian $\mathcal H$
in the sense:
\ben
{dI\over dt}= {\partial I\over\partial t} + [I,\mathcal H]_{\mathrm Poisson~bracket} = 0
\label{poisson}
\een
Therefore, in the early universe one can write down an invariant quadratic form in the Hubble 
parameter and the logarithm of the scale factor with time dependent coefficients. These coefficients 
are functions of the solutions of the Ermakov equation.

Let us now determine what type of solutions of $\rho$ are possible. Note that this is determined 
solely through (\ref{erma}) and depend on the constant 
$M={3\over 4\pi G}$ and the frequency  $\Omega(t)$ which in turn 
is determined by the scalar field $\phi$.
So a choice for $\Omega(t)$ must ensure that  a solution for $\rho$ exists and 
one also gets a scalar field consistent with cosmological scenarios.

{\it Case a}

Consider a scalar field potential $V= {\frac 1 2} m^{2}\phi^{2}$ where $m$ is the mass of the scalar 
field. If one assumes a scenario where $\dot\phi ^{2} >> V $ i.e. the kinetic energy is large compared 
to the potential energy then a solution for the scalar field is \cite{muk}
\ben
\phi(t)= const. - (12\pi)^{-1/2} ~ ln t
\label{phi2}
\een
Choosing $\Omega(t)=t^{-1/2}$ and using (\ref{phi1}) gives 
\ben
\phi(t)=\phi_{0} - {1\over 24\pi G {\sqrt C} V} ln~t
\label{phi3}
\een
where the constant of integration has been identified as $\phi_{0}$.
Comparing (\ref{phi2}) and (\ref{phi3}) will fix the constant $\phi_{0}$ and ${\sqrt C}$.
For this choice of $\Omega(t)=t^{-1/2}$ the general solution for $\rho$ is \cite{lewis}
\ben
\rho(t)=\gamma_{1}\pi M t^{1/2}\biggl[A^{2}Y_{1}^{2}(2Mt^{1/2})+ B^{2}J_{1}^{2}(2Mt^{1/2})\nonumber\\ 
+2\gamma_{2}(A^{2}B^{2}-{1\over \pi ^{2}M^{2}})^{1/2} J_{1}(2Mt^{1/2}) Y_{1}(2Mt^{1/2})\biggr]^{1/2}
\label{ro1}
\een 
Here $\gamma_{1}=\pm 1$, $\gamma_{2}=\pm 1$, $A,B$ are arbitrary complex constants, $J_{1},Y_{1}$ are
Bessel functions of the first and second kinds respectively. Let us take $\gamma_{1}=\gamma_{2}=+1$. 

Thus this choice of $\Omega$ is consistent with an ultrahard equation of state and
a scalar field with a logarithmic dependence on time \cite{muk}. Note that the dominance
of kinetic energy is a natural choice for $k-$essence scalar fields.

Now we show that the Ermakov invariant (\ref{inv2}) is a powerful tool to estimate the scale factor.
For $t\rightarrow 0$, 
$J_{\alpha}(x)\rightarrow {1\over\Gamma (\alpha +1)}({x\over 2})^{\alpha}~~;
 Y_{\alpha}(x)\rightarrow -{\Gamma (\alpha)\over\pi}({2\over x})^{\alpha}~~;0<x\leq{\sqrt (\alpha +1)};
~~\alpha > 0$.  In our case $\alpha=1$ . Using these (\ref{ro1}) takes the form  
\ben
\rho(t)= (\pi M)^{1/2}t^{3/2} + (\pi M)^{-3/2} t^{-1/2}
\label{ro2}
\een
One can now determine $A(t), B(t), C(t)$. For small times only the inverse powers of $t$ will 
dominate. Therefore keeping only $O(t^{-2})$ and $O(t^{-3})$ terms, (\ref{inv2}) becomes
($q= ln~~a(t)$):
\ben
{q\dot q\over 2\pi ^{3} M^{4} t^{2}} + {q^{2}\over 8\pi ^{3}M^{5} t^{3}}\approx I
\label{inv3}
\een
and the solution for $q$ is 
\ben 
q(t)=\biggl[A_{0} t^{-1/2M} + A_{1}t^{3}\biggr]^{1/2}
\label{q}
\een
So the scale factor is 
\ben
a(t)=e^{[A_{0} t^{-1/2M} + A_{1}t^{3}]^{1/2}}\sim e^{A_{0}^{1/2}t^{-1/4M}}
\label{solution}
\een
and the solution is consistent with the inflationary scenario. Here $A_{0}$ is an arbitrary constant 
of integration and $A_{1}= {8\pi^{3}M^{5}I\over 6M+1}$. Note that as
$a\rightarrow e^{0}=1$ , $t\rightarrow [{-2A_{0}\over A_{1}}]^{2M/(6M+1)}$. So $A_{0}$ 
should be chosen to be negative.  
(Here we have illustrated solutions for 
$(\gamma_{1}=1,\gamma_{2}=1)$. Solutions for $(\gamma_{1}=-1,\gamma_{2}=\pm 1)$ in the 
cosmological context  will be discussed in subsequent publications).

{\it Case b}

Now suppose we choose $\Omega$ to be a constant. Then (\ref{phi1}) gives 
\ben
\phi(t)=\phi_{i} - {1\over 24\pi G {\sqrt C} V} t
\label{phi4}
\een
For a constant $\Omega$ the general solution for $\rho$ is
\ben
\rho(t)=\gamma_{1}\Omega^{-1}\biggl[A^{2}cos^{2}(M\Omega t)+ B^{2}sin^{2}(M\Omega t)\nonumber\\
+2\gamma_{2}(A^{2}B^{2}-\Omega^{2})^{1/2} sin(\Omega Mt) cos(\Omega Mt)\biggr]^{1/2}
\label{solro2}
\een
Compare this with the attractor solution in \cite{muk} :
\ben
\phi_{\mathrm atr}(t)\approx \phi_{i} - {m\over {\sqrt 12\pi}}(t-t_{i})
\label{phi5}
\een
Here the trajectory joins the attractor where it is flat at $|\phi|>>1$ and afterwards the 
solution describes a stage of accelerated expansion \cite{muk}
However, $\Omega=constant$ is {\it not } a natural choice for $k-$essence fields because 
it implies that the potential energy now dominates. The discussion of this case is merely
for illustrative purposes.

The Ermakov invariant also exists in the quantum context \cite{lewis}.The invariant $I$ , now 
an operator, is a constant of motion for the quantum system for any $\rho$ that satisfies  
(\ref{erma}). So we now have the Heisenberg equation of motion for  $I$ as 
${dI\over dt}={\partial I\over\partial t}  + {1\over i\hbar}[I,H]=0$.
Creation and annihilation operators can be constructed and
normalised eigenstates of $I$ exist. Note that the hamiltonian corresponding to (\ref{lag5})
is ${\mathcal H(t)}= {p^{2}\over 2M} + {1\over 2}M\Omega^{2}(t) q^{2}$. If $\psi_{n}(q,t)$ be the 
eigenfunctions of the invariant operator $I$, then
$\langle\psi_{n}|{\mathcal H(t)}|\psi_{n}\rangle
= {M\over 2}(\rho^{-2}+\Omega^{2}\rho^{2}+{1\over M^{2}}\dot\rho ^{2})(n+{1\over 2})\hbar$
where $n=0,1,2.....$. So in a quantum context, this invariant can also be used to
estimate the quantum expectation value of a function involving the scale factor and the
Hubble parameter. An analogue of the Berry's phase in early universe can also 
be defined as follows \cite{dan}. When the time dependent parameters of a quantum system evolving 
adiabatically in time executes a complete loop in parameter space, the wavefunction 
(in addition to its dynamic phase) picks up a geometric phase. In the Ermakov context, 
this phase factor is given by 
$\gamma_{n}(\mathcal C)= -(1/2)(n+1/2)\int_{0}^{T}(\rho\ddot\rho - \dot\rho ^{2})$.
where $\rho$ satisfies (\ref{erma}). These aspects will be discussed in subsequent publications.

{\bf 3. Conclusion: An observational probe in the early universe}

The basic conclusion of this work is that the Ermakov invariant in a cosmological context 
(\ref{inv2}) can be used as an observational probe in the early universe in the domain  
{\it viz.} $-1< q(=ln~~ a(t))<1$ in the following way. Observationally or otherwise, 
the Hubble parameter and the scale factor is known over a substantially large period 
of time. If these can be known for periods within the domain under consideration, 
the validity of (\ref{inv2}) can be tested. Alternatively, knowing either  of the two  
{\it viz.} the Hubble parameter or the scale factor, 
will enable the other to be determined using (\ref{inv2}) and (\ref{functro}). As 
everything is based on a particular form of the dark energy lagrangian, 
any vindication of (\ref{inv2}) implies an indirect proof of the presence of dark energy 
as a principal constituent of the universe.

The author would like to thank the Centre For Astroparticle Physics and Space Science,
Bose Institute, Kolkata, for a sabbatical tenure during which this work has been done.

\end{document}